\documentclass[sigconf,nonacm]{acmart}
\usepackage{standalone}

\providecommand{\paperroot}{.}
\providecommand{\paperroot}{..}

\usepackage{tikz}
\usetikzlibrary{fpu}
\tikzset{>=stealth}

\newcommand{\bignum}[1]{%
	\pgfkeys{/pgf/fpu=true}%
	\pgfmathparse{#1}%
	\pgfmathprintnumber[fixed]{\pgfmathresult}%
	\pgfkeys{/pgf/fpu=false}%
}

\title{Conflicting Privacy Preference Signals in the Wild}

\begin{document}
\newcommand{\ambiguousSignalPct}{$5\%$}
\newcommand{\browserAndroidPct}{$6\%$}
\newcommand{\browserChromeheadlessPct}{$0\%$}
\newcommand{\browserChromiumPct}{$53\%$}
\newcommand{\browserEdgePct}{$4\%$}
\newcommand{\browserFirefoxPct}{$23\%$}
\newcommand{\browserIphonePct}{$3\%$}
\newcommand{\browserMobilePct}{$9\%$}
\newcommand{\browserOperaPct}{$2\%$}
\newcommand{\browserOtherPct}{$1\%$}
\newcommand{\browserSafariPct}{$8\%$}
\newcommand{\clickAcceptDntNoneChisq}{$\chi^2(1, N=1960) = 24.49, p<0.01$}
\newcommand{\clickAcceptGpcDntChisq}{$\chi^2(1, N=346) = 0.07, p=0.80$}
\newcommand{\clickAcceptGpcNoneChisq}{$\chi^2(1, N=1674) = 4.50, p=0.03$}
\newcommand{\clickAcceptPct}{$86\%$}
\newcommand{\clickAcceptPctDnt}{$77\%$}
\newcommand{\clickAcceptPctGpc}{$73\%$}
\newcommand{\dialogShownPct}{$14\%$}
\newcommand{\dialogShownPctDnt}{$11\%$}
\newcommand{\dialogShownPctGpc}{$6\%$}
\newcommand{\doesNotLoadPct}{$27\%$}
\newcommand{\doesNotLoadPctDnt}{$50\%$}
\newcommand{\doesNotLoadPctGpc}{$73\%$}
\newcommand{\moreLikelyToRejectFactorDnt}{1.9}
\newcommand{\moreLikelyToRejectFactorGpc}{2.0}
\newcommand{\uniqueImpressions}{16761}
\newcommand{\uniqueIps}{8033}
\newcommand{\uniqueSubnets}{7432}

\author{Maximilian Hils}
\email{maximilian.hils@uibk.ac.at}
\affiliation{\institution{University of Innsbruck}\country{Austria}}
\author{Daniel W. Woods}
\email{daniel.woods@uibk.ac.at}	
\affiliation{\institution{University of Innsbruck}\country{Austria}}
\author{Rainer Böhme}
\email{rainer.boehme@uibk.ac.at}
\affiliation{\institution{University of Innsbruck}\country{Austria}}

\begin{abstract}
Privacy preference signals allow users to express preferences over how their personal data is processed.
These signals become important in determining privacy outcomes when they reference an enforceable legal basis, as is the case with recent signals
such as the Global Privacy Control and the Transparency \& Consent Framework.
However, the coexistence of multiple privacy preference signals creates ambiguity as users may transmit more than one signal.
This paper collects evidence about ambiguity flowing from the aforementioned two signals and the historic Do Not Track signal.
We provide the first empirical evidence that ambiguous signals are sent by web users in the wild.
We also show that preferences stored in the browser are reliable predictors of privacy preferences expressed in web dialogs.
Finally, we provide the first evidence that popular cookie dialogs are blocked by the majority of users who adopted the Do Not Track and Global Privacy Control standards.
These empirical results inform forthcoming legal debates about how to interpret privacy preference signals.
\end{abstract}

\begin{teaserfigure}
	
\ifstandalone\begin{figure}\fi
	\centering
	\LARGE
	\setlength{\fboxsep}{8pt}
	\fbox{\begin{minipage}{15.2em}	
	\texttt{%
	GET / HTTP/1.1\\%
	Cookie:{\color{red}\ euconsent-v2=<...>}%
	\\%
	{\color{green!50!black}Sec-GPC: 1}%
	}
	\end{minipage}}
	\caption{Which privacy preference signal takes precedence\,?}
	\label{fig/conflict}
\ifstandalone\end{figure}\fi

\end{teaserfigure}

\maketitle

\section{Introduction}

Privacy laws like GDPR and CCPA empower users, at least in theory, to control how their personal data is processed. %
To do so, individuals must be able to communicate their privacy preferences with data controllers.
Web standards for privacy preference signals make this convenient and easy.
However, the literature suggests coordinating senders and recipients to adopt one standard has failed multiple times due to the competing interests of stakeholders~\cite{hils2021privacy}.
The same competing interests lead stakeholders to propose different signals, which has resulted in the coexistence of multiple signals. 
Users may transmit more than one signal and thereby express conflicting or ambiguous preferences, which creates uncertainty over which legal rules apply.

Multiple signals can be sent when signals are collected at different technical layers. 
In this study, we focus on the two dominant ways for users to express privacy preferences: on individual websites and globally in their browser.
The first approach is chosen by the Transparency \& Consent Framework (TCF), a standard developed by the Interactive Advertising Bureau and adopted by hundreds of ad-tech vendors and thousands of websites~\cite{matte2020do}.
The second approach was chosen by the Do Not Track (DNT) mechanism~\cite{mayer2011not} and also the Global Privacy Control (GPC), which now boasts over 40 million users~\cite{gpc2021fourty}.  %

The possibility of users sending multiple signals raises questions about legal interpretation under both the CCPA and the GDPR.
But before the lawyers weigh in, there are empirical questions to be answered regarding which signals users send in the wild.  
We observe $16$k impressions on websites that embed TCF dialogs and simultaneously detect the presence of a DNT and/or GPC signal.  

Our results uncover a number of sources of ambiguity not previously identified in the literature.  
First, an industry standard dialog for collecting TCF signals is blocked by \doesNotLoadPct{} of users, and this percentage rises to \doesNotLoadPctDnt{}/\doesNotLoadPctGpc{} of users with DNT/GPC turned on.  
Second, users who send a GPC signal are two %
times more likely to withhold consent than other users, which suggests the signal captures genuine privacy preferences.
Finally, even though they are more likely to not give consent, \clickAcceptPctGpc{} of users with GPC turned on still consent to being tracked by clicking ``I Accept'' in a TCF consent dialog. This shows that conflicting signals are a reality.  

Section~\ref{background} provides brief background on relevant laws and signals.
Section~\ref{methods} describes our research design.  
Section~\ref{results} presents the results.  
Section~\ref{discussion} discusses the results and suggests directions for future work.  
Section~\ref{conclusion} concludes the paper.

\section{Background} ~\label{background}
We split the background into laws and privacy preference signals.

\paragraph{Privacy Laws} The two laws most relevant to privacy preference signals are the General Data Protection Regulation (GDPR) and the California Consumer Privacy Act (CCPA).  
Article~6 of the GDPR~\cite{GDPR16} establishes a number of legal bases for processing personal data, of which (opt-in) consent is the most common legal basis claimed in a sample of hundreds of AdTech vendors~\cite[Fig.\,2]{matte2020purposes}.
Article~4 of the GDPR~\cite{GDPR16} defines consent as any freely given, specific, informed and unambiguous indication of the data subject’s wishes.

Taking a different approach, the CCPA establishes ``the right to direct a business to not sell consumers' personal information''~\cite[p.\,15]{zimmeck2020standardizing}.  
For the purposes of this paper, it is important to note that both laws link the legality of data processing to the privacy preferences of users, creating a need for signals that communicate preferences.

\paragraph{Privacy Preference Signals} \citet{hils2021privacy} identify five signals that have been adopted at various points in the last 20 years.
We ignore P3P because it was deprecated in $2017$ and ignore NAI opt-outs as they remain an unpopular and narrow signal~\cite{dixon2007network}.  
We focus on the remaining three signals.  
DNT and GPC have a similar technical design in that they extend HTTP headers by a single bit signal, but they differ in semantics.
The law does not require recipients to respect DNT, and many ad-tech companies in fact decided to ignore the signal~\cite[p.\,15]{zimmeck2020standardizing}. Nonetheless, it can still be turned on in Chrome's and Firefox's settings dialog. %
In contrast, the Global Privacy Control is designed to trigger the ``Do Not Sell'' clause (the aforementioned legal right'~\cite[p.\,15]{zimmeck2020standardizing}) under the CCPA. It also provides a possible interpretation under the GDPR in its specification. However, major browsers have not adopted GPC yet outside of browser extensions.

The third signal, the TCF, is collected via dialogs embedded in the webpage.
TCF signals can only be collected by registered intermediaries, of which QuantCast and OneTrust are the most popular~\cite{hils2021privacy}.
The semantics of this signal are much more complex~\cite{matte2020purposes, santos2020cookie} but revolve around opting-in to various data processing purposes where consent is required.
We discuss the nature of the ambiguity resulting from sending DNT/GPC opt-outs and TCF opt-ins in Section~\ref{discussion}.

Finally, it is worth noting that niche and emerging signals exist that we did not consider.
Do Not Sell signals can also be collected via webpages and stored as cookies, which use the standardized \emph{US Privacy String} format \cite{iab2020usprivacystring}.
These cookies were successfully reset using the OptMeowt add-on for $17$ of $30$ websites in a recent study~\cite{zimmeck2020standardizing}.  A technical specification for the Advanced Data Protection Control~\cite{adpc} was proposed that could automatically send privacy preference signals including TCF and Do Not Sell, but does not define any new signals in terms of semantics.

\section{Method} \label{methods}

To examine the interplay between privacy preference signals,
we embedded Quantcast's cookie consent dialog on the landing page of three websites for a short period of time and also logged visitors' DNT/GPC headers.
In constrast to previous research~\cite{hils2020measuring}, we not only measure a user's decision when they are presented with a TCF consent dialog, but also if they are shown a consent dialog at all. This is important as our findings indicate that a non-negligible number of users employ techniques that block popular consent dialogs entirely. 

\paragraph{Study Participants}
We sampled a very technical audience on all three websites.
The majority of our measurements were made on \href{https://mitmproxy.org/}{mitmproxy.org}, the website of an open-source program primarily used by software developers (72\% of all impressions) \cite{mitmproxy}. Additionally, our research group's website and the website of a Capture The Flag contest we hosted contributed 14\% of observations each. 
Note that all numbers in this paper are reported as impressions of the landing page. We do not perform any additional grouping to not overrepresent users who employ additional privacy measures (such as clearing cookies). In total, we observe \bignum\uniqueImpressions{} impressions by \bignum\uniqueIps{} IPv4 addresses from \bignum\uniqueSubnets{} /24 subnets.

In terms of browsers used,
\browserChromiumPct{} of visitors used Chrome/Chromium, \browserFirefoxPct{} Firefox, \browserMobilePct{} mobile browsers, \browserSafariPct{} Safari, and \browserEdgePct{} Edge. For comparison, Wikimedia reports 55\% Chrome, 13\% Firefox, 10\% Safari, and 8\% Edge on their desktop sites (June 2021).

	\newcommand{\iIncomplete}{2004}
\newcommand{\iBlocked}{4442}
\newcommand{\iConsentAccept}{1689}
\newcommand{\iConsentCurrent}{3357}
\newcommand{\iConsentReject}{271}
\newcommand{\iDialogLeftOpen}{322}
\newcommand{\iNotEu}{6680}
\newcommand{\iGpcTrueBlocked}{402}
\newcommand{\iGpcTrueConsentAccept}{22}
\newcommand{\iGpcTrueConsentCurrent}{43}
\newcommand{\iGpcTrueConsentReject}{8}
\newcommand{\iGpcTrueDialogLeftOpen}{4}
\newcommand{\iGpcTrueNotEu}{72}
\newcommand{\iGpcFalseBlocked}{4040}
\newcommand{\iGpcFalseConsentAccept}{1667}
\newcommand{\iGpcFalseConsentCurrent}{3314}
\newcommand{\iGpcFalseConsentReject}{263}
\newcommand{\iGpcFalseDialogLeftOpen}{318}
\newcommand{\iGpcFalseNotEu}{6608}
\newcommand{\iDntTrueBlocked}{1760}
\newcommand{\iDntTrueConsentAccept}{244}
\newcommand{\iDntTrueConsentCurrent}{442}
\newcommand{\iDntTrueConsentReject}{72}
\newcommand{\iDntTrueDialogLeftOpen}{80}
\newcommand{\iDntTrueNotEu}{895}
\newcommand{\iDntFalseBlocked}{2682}
\newcommand{\iDntFalseConsentAccept}{1445}
\newcommand{\iDntFalseConsentCurrent}{2915}
\newcommand{\iDntFalseConsentReject}{199}
\newcommand{\iDntFalseDialogLeftOpen}{242}
\newcommand{\iDntFalseNotEu}{5785}

	\begin{figure}
		\begin{tikzpicture}[y=1cm,x=1cm]
			\node at (0,0) {user visits page};
			\node at (0,-1) {browser blocks dialog?};
			\node at (0,-2) {GDPR applies?};
			\node at (0,-3) {exist. cookie?};
			\node at (0,-4) {decision?};
			
			\begin{scope}[shorten >=0.2cm,shorten <=.25cm,->]
				\pgfkeys{/pgf/fpu}
				\pgfmathsetmacro{\decision}{\iConsentAccept+\iConsentReject+\iDialogLeftOpen}
				\pgfmathsetmacro{\hascookie}{\decision+\iConsentCurrent}
				\pgfmathsetmacro{\gdprapplies}{\hascookie+\iNotEu}
				\pgfmathsetmacro{\blocking}{\gdprapplies+\iBlocked}
				\pgfmathsetmacro{\total}{\blocking+\iIncomplete}
				\pgfkeys{/pgf/fpu=false}
				
				\draw (0,-0) -- (0,-1) node [midway, right] {\bignum\blocking};
				\draw (0,-1) -- (0,-2) node [midway, right] {\bignum\gdprapplies};
				\draw (0,-2) -- (0,-3) node [midway, right] {\bignum\hascookie};
				\draw (0,-3) -- (0,-4) node [midway, right] {\bignum\decision};
				
				\pgfkeys{/pgf/fpu=false}

				\draw (1.6,-0) -- (3.1,-0) node[midway,below] {\bignum\iIncomplete} node[right=-.1cm, text width=2.1cm] {incomplete\\measurement};
				\draw (1.6,-1) -- (3.1,-1) node[midway,below] {\bignum\iBlocked} node[right=-.1cm, text width=2cm] {request\\blocked};
				\draw (1.6,-2) -- (3.1,-2) node[midway,below] {\bignum\iNotEu} node[right=-.1cm, text width=2cm] {outside EU};
				\draw (1.6,-3) -- (3.1,-3) node[midway,below] {\bignum\iConsentCurrent} node[right=-.1cm, text width=2.2cm] {decision\,reused\\from\,prev.\,visit};
				\draw (1.6,-4) -- (3.1,-4) node[midway,below] {\bignum\iDialogLeftOpen} node[right=-.1cm, text width=2.1cm] {shown but no\\interaction};
				
				\draw (-.5,-4) -- (-.7,-5) node[midway, left] {\bignum\iConsentAccept} node[below=-.2cm] {accept\strut};
				\draw (.5,-4) -- (.7,-5) node[midway, right] {\bignum\iConsentReject} node[below=-.2cm] {reject\strut};
					
		\end{scope}
			
		\end{tikzpicture}
		\caption{Our Consent Dialog Measurement Pipeline}
		\label{fig/funnel}
	\end{figure}
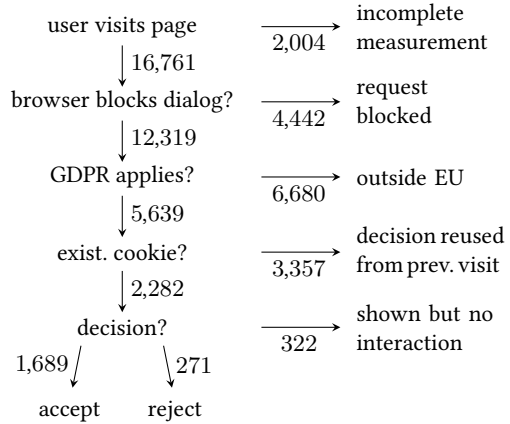

\paragraph{Data Collection}
During our study period we embedded a logging script on all three websites and recorded the following data (see Figure \ref{fig/funnel}):

\begin{enumerate}
	\item the user's browser and whether they sent a GPC/DNT header;
	\item state transitions from the browser's page visibility API;
	\item whether Quantcast's dialog could be loaded.
\end{enumerate}
If loading the dialog was successful, we additionally recorded:
\begin{enumerate}
	\setcounter{enumi}{3}
	\item Quantcast's assessment of whether GDPR applies to the current user;
	\item existing consent decisions from previous visits;
	\item the user's (new) consent decision when a dialog is shown.
\end{enumerate}
To determine whether Quantcast's dialog can be loaded, we manually inject its main script tag and add \texttt{load} and \texttt{error} event listeners. We cannot detect what prevents this resource from loading, but we suspect that the majority is via ad-blocking browser extensions and DNS-level content blockers. 

If Quantcast's JavaScript code could be loaded successfully, we interact with their implementation of the TCF API to determine if GDPR applies to the current user. 
We only show dialogs to users in the EU, which is Quantcast's default setting.

\begin{figure}[h]
		\fbox{
			\includegraphics[width=.95\linewidth]{\paperroot/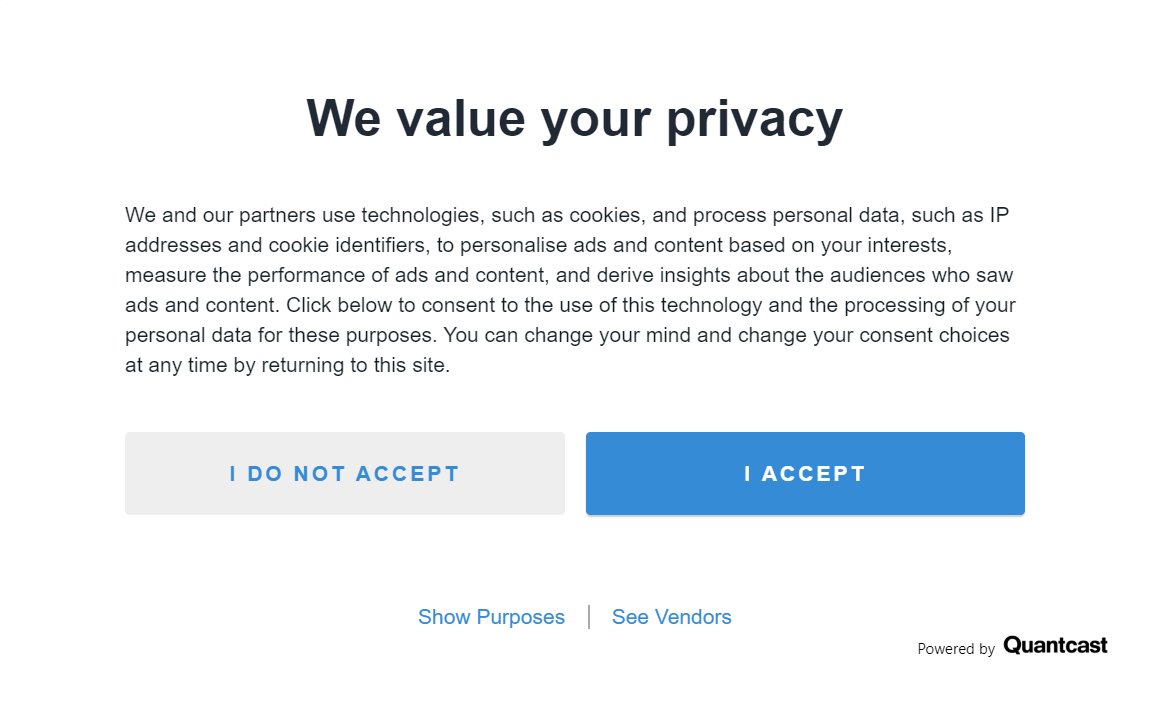}
		}
		\caption{Quantcast's TCF cookie dialog used in this study. Note that we used Quantcast's default configuration, which renders the ``I accept'' button more prominently and thus does not provide an equal choice.}
		\label{fig/qcdialog}
	\end{figure}

If we find that GDPR applies, we present the user with a dialog (see Figure~\ref{fig/qcdialog}) unless a decision has already been made.
We track prior decisions via a cookie set by the dialog.  
This means the individuals behind the decisions are unique to the extent these cookies are preserved. 
In total, a consent dialog is only shown for \dialogShownPct{} of impressions.

To get a more accurate picture of the user's interaction with the web page, we observe the browser's page visibility API, which emits \texttt{visibilitychange} events when the page becomes visible, hidden, or closed. We use the Beacon API (\texttt{navigator.sendBeacon}) to record all events as this interface still works when the page is closed. As of 2021, both APIs are available in all major browsers. We discard all measurements for which we did not receive a complete set of events. This also reduces the impact of general network problems, which otherwise may be misattributed as blocking.

\paragraph{Research Ethics}
Our research design requires ethical consideration as each dialog takes user time.
Given thousands of websites already impose these dialogs on users~\cite{matte2020do}, we judge that the time cost is outweighed by the value of information derived from our study in shaping the design and adoption of these dialogs.
Previous research has shown that Quantcast's dialogs are completed in 3.2s on average~\cite{hils2020measuring}.
While our institution does not require IRB review for minimal risk studies, we ensured that we did not deceive or harm website visitors and their privacy. All displayed consent notices functioned as described and respected the visitor's choice.\\
Explaining the research purpose before/after the experiment would lead to a much longer interruption than the initial dialog.

\section{Results} \label{results}
We split our results into two aspects of ambiguity, namely blocked dialogs and multiple signals, and then consider robustness.

\begin{figure}
	\centering
	
	\definecolor{blocked-color}{HTML}{E69F00}
	\definecolor{shown-color}{HTML}{ffffff}
	\definecolor{notshown-color}{HTML}{cccccc}
	
	\definecolor{consent-color}{HTML}{009E73}
	\definecolor{reject-color}{HTML}{D55E00}

	\begin{tikzpicture}[y=.75cm, x=6.8cm,
		blocked/.style={fill=blocked-color},
		notshown/.style={fill=notshown-color},
		shown/.style={fill=shown-color},
		consent/.style={fill=consent-color},
		reject/.style={fill=reject-color},
	]
	
		\newcommand{\gpcplot}[2]{
			
			\node[rotate=90, inner sep=0] at (-.3cm, -.55) {#2};
		
			\pgfmathsetmacro{\notshown}{
				\csname i#1NotEu\endcsname +
				\csname i#1ConsentCurrent\endcsname +
				\csname i#1DialogLeftOpen\endcsname
			}
			\pgfmathsetmacro{\shown}{
				\csname i#1ConsentAccept\endcsname +
				\csname i#1ConsentReject\endcsname
			}
			\pgfmathsetmacro{\total}{
				\csname i#1Blocked\endcsname +
				\shown +
				\notshown
			}
		
			\pgfmathsetmacro{\a}{(\shown + \notshown) / \total}
			\pgfmathsetmacro{\b}{\notshown / \total}
			
			\begin{scope}
				\clip(0,0) rectangle (1,-1);
				\shade[draw=darkgray, 
					very thin,
					top color=darkgray!5!white, 
					bottom color=darkgray!25!white
				] (\b,-0.42) -- (\a,-.42) -- (1,-.9) -- ++(-1, 0) -- cycle;
			\end{scope}

			\path[blocked,draw=darkgray] (0,0)++(0,-0.42)  rectangle ++(1,.84);
			\path[shown,draw=darkgray] (0,0)++(0,-0.42)  rectangle ++(\a,.84);
			\path[notshown,draw=darkgray] (0,0)++(0,-0.42)  rectangle ++(\b,.84);
			
			\node[right, text width=.3cm] at (1,0) {\scriptsize $N$\texttt{=}$\bignum\total$};
			
			\begin{scope}[shift={(0,-.9)}]
				\acceptreject{#1}
			\end{scope}
		}
	
		\newcommand{\acceptreject}[1]{
			\pgfmathsetmacro{\shown}{
				\csname i#1ConsentAccept\endcsname +
				\csname i#1ConsentReject\endcsname
			}
			\pgfmathsetmacro{\accept}{
				\csname i#1ConsentAccept\endcsname / \shown
			}
			\path[reject,draw=darkgray] (0,0)  rectangle ++(1,-.5);
			\path[consent,draw=darkgray] (0,0)  rectangle ++(\accept,-.5);
			\node[right] at (1,-.25) {\scriptsize $N$\texttt{=}$\bignum\shown$};
		}
	
		\begin{scope}[shift={(0,1.5)}]
			\gpcplot{GpcTrue}{GPC}
		\end{scope}
		\begin{scope}[shift={(0,3.8)}]
			\gpcplot{DntTrue}{DNT}
		\end{scope}
		\begin{scope}[shift={(0,6.1)}]
			\gpcplot{DntFalse}{none}
		\end{scope}

		\begin{scope}[yshift=-.06cm]
			\draw (0,0) -- ++(1,0);
			\draw (0,0) -- ++(0,-3pt) node[below right=-1pt] {0\%};
			\draw (1,0) -- ++(0,-3pt) node[below left=-1pt] {100\%};
			\foreach \i in {.1,.2,.3,.4,.5,.6,.7,.8,.9} {
				\draw (\i,0) -- ++(0,-2pt);	
			}
		\end{scope}
		
		\node at (.5,-.5cm) {\% of users};

		\node[right, inner sep=0] at (0,-1cm) {Dialog:};
		\node[right, inner sep=0] at (0,-1.5cm) {Signal:};

		\foreach \x/\y/\c/\lbl in {
			.2/-1cm/notshown/not shown,
			.525/-1cm/shown/shown,
			.75/-1cm/blocked/blocked,
			.2/-1.5cm/consent/accept,
			.43/-1.5cm/reject/reject
		} {
			\draw[\c, draw=darkgray] (\x, \y) ++ (0,-.5ex) rectangle ++(1.5ex,1.5ex) -- ++(0,-.95ex) node[right] {\strut \lbl};
		}
		
	\end{tikzpicture}\vspace{-1ex}
	\caption{
		The majority of users who have GPC turned on (bottom pair of bars) outright block consent dialogs. 
		Those who do not block are relatively more likely to reject tracking when presented with a dialog offering equal choice. 
		Still, the majority of GPC users click ``Accept'' and thus send an ambiguous signal.
		The top pair of bars show the baseline without browser-based privacy signal.
		The deprecated DNT signal (middle) was more prevalent than GPC, exhibited less extreme blocking, and a similar behavioural response.
	}
	\label{fig/gpc}
\end{figure}
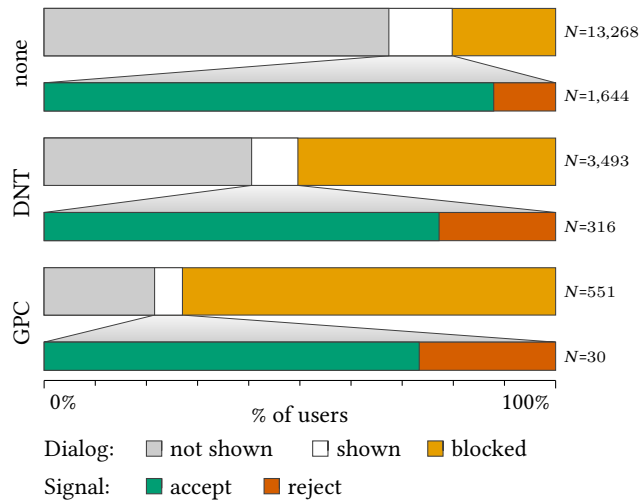

\paragraph{Blocked Dialogs}
The top bars in Figure~\ref{fig/gpc} show that collecting a TCF signal via a dialog is non-trivial.
The default version of the market-leading dialog does not load for \doesNotLoadPct{} of users, which rises to \doesNotLoadPctDnt{}/\doesNotLoadPctGpc{} of users with DNT/GPC enabled.
This technical response---blocking the privacy preference communication channel---was not previously considered in the literature.  

\paragraph{Multiple Signals}
The second source of ambiguity results from users sending multiple signals simultaneously.
First, $3.5$k/$550$ of impressions send a DNT/GPC signal respectively. 
All but three of the GPC impressions also send a DNT signal so we do not further differentiate.
Displaying a TCF dialog to these impressions creates the potential for a conflict, namely when users send an accept TCF signal while also sending a DNT/GPC signal.
Such a conflict occured for \ambiguousSignalPct{} of all impressions, or \clickAcceptPctDnt{}/\clickAcceptPctGpc{} when looking at DNT/GPC-enabled users only.
We discuss the nature of the ambiguity later in the paper.%

The co-existence of these signals does not only increase ambiguity as the DNT/GPC signals have explanatory power over privacy preferences expressed via the TCF signal.
Users with DNT/GPC enabled were \moreLikelyToRejectFactorDnt{}/\moreLikelyToRejectFactorGpc{} times more likely to click ``I do not accept'' on the TCF dialog than those without. These results are significant at the $p < 0.01$ level for DNT and $p < 0.05$ for GPC.
Note we cannot reject the null hypothesis that DNT/GPC are drawn from the same distribution (\clickAcceptGpcDntChisq{}).

\paragraph{Robustness}
We run a number of checks to reduce the risk of spurious findings. 
It could be that GPC adoption was driven by browsers and browser extensions turning it on by default, as it was done by Brave browser~\cite{brave2021gpc}.
In our sample, Firefox users are most likely to send a GPC signal (8.9\,\%), followed by Chromium (2.3\,\%) and Edge (1.6\,\%).
The share of GPC signals from other user agents is statistically zero.
Note that Chromium includes Chrome, the most popular browser on the web, as well as Brave, a niche browser catering pro-privacy and cryptocoin-savy users, which identifies as Chrome in the user-agent string.
To our knowledge, at the time of our study Brave was the only browser that sent GPC signals by default without asking the user.
While we suspect that a number of the Chromium cases with GPC turned on do originate from Brave, the fact that other browsers emit more GPC signals, in both relative and absolute numbers, reassures us that the results are not purely driven by a single browser's default setting.
The finding that users who emit DNT or GPC signals tend to chose more privacy-minded TCF options corroborates the behavioral interpretation.

\section{Discussion} \label{discussion}
We discuss how to collect preferences, the nature of the ambiguity,  and the validity of our results.

\paragraph{Collecting Preferences} Research into GDPR cookie consent dialogs consistently shows that dialogs contain dark patterns that erode user autonomy~\cite{boehme2010, adjerid2013sleights, utz2019uninformed, mathur2019dark, machuletz2020, nouwens2020dark, oconnor2020unclear, habib2020scavenger}.
Our findings could be interpreted as further evidence that industry standard dialogs lead users to express untrue preferences.  For example, \clickAcceptPctDnt{} of DNT-enabled users accept data processing in a TCF dialog despite sending a global ``Do Not Track'' signal.

On the other hand, one could argue that the browser-controlled signals do not capture the true preferences of users. 
For example, Brave browser turns the signal on by default and does not provide an off-toggle~\cite{brave2021gpc}. 
The lack of an off-toggle goes even further than Microsoft's decision to turn DNT on by default in 2012, which led an AdTech industry group to withdraw from the DNT initiative~\cite{iab2012dnt}.
Notably, new counter-arguments exist given that the CCPA establishes that ``affirmatively choosing products or services with privacy-protective features\dots is considered a sufficiently clear manifestation of opting out''~\cite{zimmeck2020standardizing}. This argument may not apply in other jurisdictions, especially given some users may adopt Brave for features beyond privacy protection (e.g.,\,cryptotokens).

Regardless of how the GPC signal is set, we have shown it explains a significant portion of the variance in expressed privacy preferences, which motivates further research into permanently storing privacy preferences in the browser.
In particular, browsers could think about collecting more than $1$\,bit from users ($0$\,bits in the case of Brave) given this information must apply across a range of jurisdictions.
Taking Colorado's new privacy law as an example, the $1$\,bit GPC signal has to cover not only the ``Do Not Sell'' clause but also opt-in consent for storing sensitive data.
While arguments can be made for how to interpret GPC's single bit under each law, the ambiguity could be used by AdTech firms to interpret the signal in their own interest or even request signals to `clarify' the situtation, which imposes yet more decision burden on users.

Our findings suggest an alternative way forward, at least for opt-in consent.
Unlike opt-out signals which default to tracking, opt-in requirements force data processors to collect a privacy preference signal.
If browsers block the interface collecting preferences, such as dialogs embedded in web-pages, then firms have no legal basis under an opt-in requirement.
This is particularly relevant given a back-of-the-envelope calculation reveals users have already wasted at least $2,500$ years~\cite{hils2021privacy} sending TCF signals.
This will likely invoke counter-measures from websites leading to an arms race~\cite{nithyanand2016adblocking}.

\paragraph{Ambiguous Signals}
Two sources of ambiguity we identified, namely blocking dialogs and multiple signals, are particularly relevant for the GDPR where many data processors~\cite{hils2020measuring} rely on an ``unambiguous''~[Art.\,4]\cite{GDPR16} opt-in consent as a legal basis.  
Blocked dialogs should be resolved in favour of the user not having provided consent.  
Multiple signals are less easily resolved and may require focusing on the semantics of each signal.
For example, the DNT signal ``represents a superset of what is covered by Do Not Sell''~\cite[p.\,15]{zimmeck2020standardizing} and so is a less ambiguous objection to some of the TCF's data processing purposes.

\sloppy
While many resolution approaches are conceivable (e.g., most/least privacy-minded, temporal or normative order, user intervention), we defer this question to legal analysis with the Blogosphere providing preliminary arguments~\cite{berjon2021gpc, pandit2021gpc}. %
The same sources of ambiguity will likely be resolved differently in non-EU jurisdictions.
This creates a technical problem for firms processing data from users across multiple jurisdictions---firms must infer each user's jurisdiction.
Could privacy aware users masquerade as residents of the jurisdiction with the strongest privacy rights?

Going beyond ambiguity, users can send two conflicting signals expressed under the same standard when both the webpage and the browser collect preferences.
For example, this could occur with the proposed ADPC signal \cite{adpc}
if a user clicked ``I accept'' in the TCF consent dialog while sending a ``No consent under TCF'' ADPC header.

\paragraph{Validity}  We argue that our study has high ecological validity in that we chose the most popular dialog~\cite{hils2021privacy} and displayed it to users browsing a real website.  
Unfortunately, we could only present this design to the users of three websites, all of which likely over-sample privacy aware and technically literate users.
The majority of our participants visited the website of an open-source program targeting developers, and the remaining from our research group's webpages.
Future work could embed the same study in a broader range of websites.
In fact, our `experiment' could be carried out passively by any website collecting TCF signals.

\section{Conclusion} \label{conclusion}
We present first evidence that websites do receive ambiguous privacy signals, namely opt-in TCF signals sent alongside a GPC opt-out signal.
Moreover, the share of ambiguous privacy signals due to blocked TCF dialogs is significant.
Both phenomena have been overlooked in the empirical literature. 
Finally, our study suggests that user adoption of the GPC helps to explain privacy preferences, and is associated with a greater propensity to reject consent.%

\bibliographystyle{ACM-Reference-Format}
\bibliography{paper}

\section*{Appendix}

\end{document}